\documentclass[pra,aps,groupaddress,prd,twocolumn]{revtex4}
\usepackage{epsfig,amsmath,graphicx,amssymb}
\def\be{\begin{equation}}
\def\ee{\end{equation}}
\def\bea{\begin{eqnarray}}
\def\eea{\end{eqnarray}}
\def\bes{\begin{subequations}}
\def\ees{\end{subequations}}
\def\bwt{\begin{widetext}}
\def\ewt{\end{widetext}}
\begin{document}
%%%%%%%%%%%%%%%%%%%%%%%%%%%%%%%%%%%%%%%%%%%%%%%%%%%%%%%%%%%%%%%%%%%%%%%%%%%%%%%%%
\title{Fourier mode dynamics for the nonlinear
Schr\"{o}dinger equation in one-dimensional bounded domains}

\author{J. G. Caputo$^{1}$, N. K. Efremidis$^{2}$, Chao Hang$^{3,4}$}
%, V. V. Konotop$^{1,3}$}
\affiliation{$^1$Laboratoire de Math\'{e}matiques, INSA de Rouen, B.
P.8, 76801 Saint-Etienne du Rouvray, France\\
E-mail: caputo@insa-rouen.fr\\
$^2$Department of applied
Mathematics, University of Crete, 71409
Heraklion, Crete, Greece\\
E-mail: nefrem@tem.uoc.gr\\
$^3$Centro de F\'isica Te\'orica e Computacional, Faculdade de
Ci\^{e}ncias, Universidade de Lisboa, Avenida Professor Gama Pinto
2, Lisboa 1649-003, Portugal\\
E-mail: chang@cii.fc.ul.pt \\
$^4$Department of Physics and State Key Laboratory of Precision
Spectroscopy, \\East China Normal University, \\
Shanghai 200062, China}

\date{\today}

%%%%%%%%%%%%%%%%%%%%%%%%%%%%%%%%%%%

\begin{abstract}

We analyze the 1D focusing nonlinear Schr\"{o}dinger equation in a finite interval
with homogeneous Dirichlet or Neumann boundary conditions. There are two main
dynamics, the collapse which is very fast and a slow cascade of Fourier modes.
For the cubic nonlinearity
the calculations show no long term energy exchange between Fourier modes
as opposed to higher nonlinearities. This slow dynamics 
is explained by fairly simple amplitude equations for the resonant
Fourier modes. Their solutions are well behaved so filtering high
frequencies prevents collapse. Finally these equations
elucidate the unique role of the zero mode for the
Neumann boundary conditions.

%Our results:\\
%(i) We find conserved quantities and virial identity for Dirichlet and Neumann.
%For Dirichlet we recover that $H<0$ is a sufficient condition for collapse, for
%$d\ge 2$. For Neumann it is more complicated, 
%$H<0$ for $d=2,3$ may not be sufficient for collapse. \\
%(ii) We find resonant mode equations, analytically. The behavior of
%periodic cascade of Fourier modes can be explained by the solutions of mode equations. \\
%(iii) From a numerical point of view, we show that filtering
%high Fourier modes prevents collapse. \\
%{\bf Things to improve or fix} It would be nice to compare runs with 
%estimates of time to collapse. Also the influence of momentum
%has been seen but it is not easy to quantify. Finally I dont understand the
%wiggles on the numerical results, they should be checked!
\end{abstract}

%\pacs{42.65.Tg, 05.45.Yv, 42.50.Gy}

\maketitle

%%%%%%%%%%%%%%%%%%%%%%%%%%%%%%%%%%%%%%%%%%%%%%%%%%%%%%%%%%%%%%%%%

\section{Introduction}

The nonlinear Schr\"{o}dinger equation in 2 or 3 spatial dimensions 
with a focusing nonlinearity has been studied intensively 
because of the collapse
phenomenon, see for example the reviews of Berge \cite{Berge} and Sulem
\cite{Sulem}. Most of these
studies have been done for an infinite domain, for which the equation is
invariant by a scaling transformation. This symmetry is important to
determine the conditions for collapse. 
In many applications like for example a laser propagating in an
optical fiber, the domain is finite so the boundaries play an important
role. A pioneering study was conducted by Fibich and Merle \cite{FM}
for the 2D cubic nonlinear Schr\"{o}dinger equation. They showed that 
circularly symmetric ground state waveguide solutions are stable 
and that the critical power condition for collapse is sharp unlike for
an infinite domain. For small amplitudes the ground states reduce to
the Bessel linear modes of the Laplacian. An interesting problem is then
how these modes exchange energy as the solution evolves. This issue is
important since some of these modes can be filtered out. More generally this energy exchange between linear modes is related to
the old problem studied by Fermi-Pasta and Ulam (see the first section
of \cite{bcs07} for subsequent developments). For a chain of oscillators with cubic nonlinearities, a medium amplitude Fourier mode initial 
condition gives rise to a cascade of higher Fourier modes and 
energy flows back into the initial mode. For the cubic nonlinear Schr\"{o}dinger
equation in 1D we expect a similar recurrence because the underlying
long wave reduction model remains the Korteweg-de Vries equation. 
This recurrence is different from what 
happens in turbulence where there is a one way flow of energy in wave
numbers. For example Muraki \cite{Muraki} studied  this one-way cascade
for the Burgers equation. Using the Cole-Hopf transformation, he was able to
quantify the phenomenon. Note finally the study \cite{zdp04} by Zakharov et al
of wave turbulence carried out on a 1D model in Fourier space.

Following a similar approach, in this article we have analyzed the
1D nonlinear Schr\"{o}dinger equation with cubic and quintic nonlinearities
on a finite interval with Dirichlet or Neumann boundary conditions.
We have chosen to work in 1D to benefit from the Fourier machinery
which enables to solve the problem fairly easily. A relatively small
number of Fourier modes are necessary to describe well the solution
when it is not singular. Another advantage is that the analysis can be
done easier than for the 2D case. We have studied the stationary solutions and
their stability. We have also obtained simple models for the resonant
transfer of energy between Fourier modes for the cubic and quintic
nonlinearities. As expected there is no transfer for the cubic
nonlinearity reflecting the integrability of the equation. For a quintic
nonlinearity a resonant transfer exists. The solutions of these reduced
models compare well to the numerical solutions of the partial differential
equation. This method of analysis of the resonant transfers of energy can
be extended to higher dimensions. The principle remains the same but of course
the machinery will be much more complicated. From another point of view
our study presents the time evolution of a solution prior to 
collapse. Finally we show that filtering prevents collapse.\\
The article is organized as follows. In section II we review the 
conservation laws and the Virial identity, the main theoretical tool,
for both the Dirichlet and the Neumann boundary conditions. Section
III presents the ground states and their stability. Numerical results
are shown in section IV and explained using a new model of resonant
energy transfer in section V. We conclude in section VI.

\section{Conserved quantities and Virial relations}

We consider the one dimensional nonlinear Schr\"{o}dinger (NLS)
equation
\be\label{nls} i\psi_t+\psi_{xx}+|\psi|^{2d}\psi=0 \ee
on a smooth, bounded domain $[0,\pi]$ with the homogeneous Dirichlet
boundary condition
\be\label{bc-d} \psi(x=0)=\psi(x=\pi)=0, \ee
or the Neumann boundary condition
\be\label{bc-n} \psi_x(x=0)=\psi_x(x=\pi)=0. \ee
Here $d$ is a positive integer with $d=1$, 2, and 3 corresponding to
the cubic, quintic, and septic nonlinearities.

Eq. (\ref{nls}) with both Dirichlet and Neumann boundary conditions
admits the following conserved quantities, the L$^2$ norm
\be\label{P} P=\int_0^{\pi}|\psi|^2 dx, \ee
which is the total power in optics and the Hamiltonian
\be\label{H}
H=\int_0^{\pi}\left(|\psi_x|^2-\frac{1}{d+1}|\psi|^{2d+2}\right) dx.
\ee
The momentum
\be\label{Pi}
\Pi=i\int_0^{\pi}(\psi\psi^{\ast}_x-\psi^{\ast}\psi_x)dx, \ee
which is conserved for the infinite domain now has a flux
\be\label{flux-d} \Pi_t=- 4 [|\psi_x|^2]_0^\pi, \ee
for Dirichlet boundary condition and
\be\label{flux-n}
\Pi_t=\left[(|\psi|^2)_{xx}+\frac{2d}{d+1}|\psi|^{2d+2}\right]_0^\pi,
\ee
for Neumann boundary condition.

We can analyze the evolution of following integral quantities
related with the model (\ref{nls})
\bes
\bea
I_1(t)&=&\int_0^{\pi}|\psi|^2x^2 dx,\\
I_2(t)&=&\int_0^{\pi}\left(|\psi_x|^2-\frac{d}{2(d+1)}|\psi|^{2d+2}\right)
dx\nonumber\\
&=&H-\frac{1}{2}\int_0^{\pi}\frac{d-2}{d+1}|\psi|^{2d+2} dx. \eea
\ees
Here, $I_1$ is the variance, which is a common tool for predicting
collapse of NLS equation solutions in infinite domain and we assume
that $I_j$ ($j=1,2$) are initially well defined.

For Dirichlet boundary, some algebra leads to
\be \label{moment-d} \frac{d^2 I_1}{dt^2}=-4\pi
[|\psi_x|^{2}]_{x=\pi} + 8H
-4\int_0^{\pi}\frac{d-2}{d+1}|\psi|^{2d+2} dx, \ee
which shows that $H<0$ is a sufficient condition for collapse. 
Notice that the right hand side of (\ref{moment-d}) can be negative
even though $H>0$. In the infinite domain the first term is absent.
Therefore, compared with the infinite line, Dirichlet boundaries 
focus the solution and enhance the collapse. 

For the Neumann boundary conditions, we have
\bea \label{moment-n} \frac{d^2
I_1}{dt^2}&=&2\pi\left[(|\psi|^2)_{xx}+\frac{2d}{d+1}|\psi|^{2d+2}\right]_{x=\pi}\nonumber\\
&+& 8H -4\int_0^{\pi}\frac{d-2}{d+1}|\psi|^{2d+2} dx. \eea
From Eq. (\ref{moment-n}) we notice that $H<0$ is not a sufficient
condition for the collapse. 
Therefore the Neumann boundary can be either reflecting or absorbing 
and enhance or suppress the collapse.

\section{Bound states}

The time periodic solutions of Eq. (\ref{nls}) can be
searched in the form
\be \psi(z,x)=u(x)\exp(iEt), \ee
where $u(x)$ is a real function and $E$ is the propagation
constant which is also real. The resulting equation then reads
\be \label{stationary} -Eu+u''+ u^{2d+1}=0 \ee
with the Dirichlet boundary conditions
\be \label{bc1-d} u(x=0)=u(x=\pi)=0 \ee
and the Neumann boundary conditions
\be \label{bc1-n} u_x(x=0)=u_x(x=\pi)=0. \ee

A conserved quantity of Eq. (\ref{stationary}) can be
found by multiplying with $u'(x)$ and integrating over $x$
\be \label{K}
K=\frac{1}{2}(u')^2-\frac{E}{2}u^2+\frac{1}{2d+2}u^{2d+2}, \ee
which can be further integrated to give
\be \int dx=\pm\frac{1}{2}\int dz
\left[z\left(4K+2Ez-\frac{2}{d+1}z^{d+1}\right)\right]^{-1/2} \ee
by defining $z=u^2$. The above integral can be expressed in terms of
Jacobi elliptic functions in both the cases of cubic and quintic
nonlinearities. However, in the latter case the resulting
expressions are rather complicated.

\begin{figure}
\centering
\includegraphics[scale=0.3]{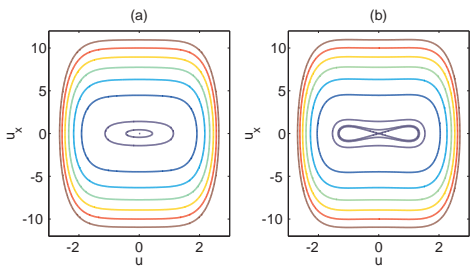}
\caption{\footnotesize Phase portrait associated with Eq. (\ref{K})
for $E=-1$ (left panel) and $E=1$ (right panel) for $d=2$. The
levels presented for $K$ are
$K=0,\,0.1,\,1,\,10,\,20,\,30,\,40,\,50,\,60$. \label{fig1}}
\end{figure}

The phase portrait associated with Eq. (\ref{K}) is shown in
Fig.~\ref{fig1} for $E=-1$ (left panel) and $E=1$ (right panel). The
points at $u=0$ ($u_x=0$) correspond to the solutions on the
boundaries for Dirichlet (Neumann) case.
%
%===========================fig1===============================%
%===========================fig1===============================%
%

If $u\ll1$, the nonlinearity does not play an important role and the
solutions are close to the linear limit of Eq. (\ref{stationary}),
i.e.
\be \label{linsol-d} u_m(x)=A\sin(mx) ,\ee
for Dirichlet boundary condition and
\be \label{linsol-n} u_m(x)=A\cos(mx) , \ee
for Neumann boundary condition. Here $m$ denotes the index of the
wave number $m=0,1,2,\cdots$ and $A$ denotes the amplitude with
$A\ll1$. The propagation constant, $K$, and the power are then given
by
\be \label{E_P} E_m=-m^2,\quad K_{m}=\frac{m^2}{2}A^2, \quad
P=\frac{\pi}{2}A^2. \ee
When $|u|$ starts to increase, the propagation constant, as well as
the form of the solutions, is slightly modified due to the effect of
nonlinearity, i.e. $E=E_m+\delta E$ with $\delta E\ll E_m$. The
relationship between the correction of the propagation constant
$\delta E$ and amplitude $A$ (and thus $P$) can be computed by
\be \label{E_P2} \delta E=
\frac{(2d+2)!}{2^{2d+1}[(d+1)!]^2}A^{2d}=\frac{(2d+2)!}{\pi^d
2^{d+1}[(d+1)!]^2}P^{d}. \ee
When $u\sim 1$, the nonlinearity has a strong effect and we have to
resort to numerical methods to find the solution.
We also notice that Eq. (\ref{stationary}) with the Neumann boundary
condition admits constant solution, i.e. $u=A_0$ with
$E=A_0^{2d}\geq0$.

\begin{figure}
\centering
\includegraphics[scale=0.35]{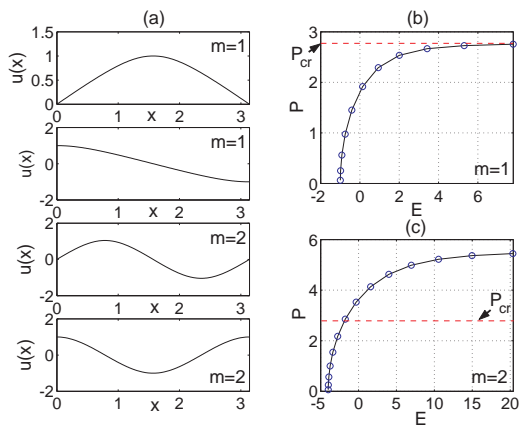}
\caption{\footnotesize The stationary solutions $u(x)$ for $d=2$.
The initial conditions are taken as $\sin(mx)$ and $\cos(mx)$
naturally satisfying the Dirichlet and Neumann boundary conditions,
respectively, with $m=1$ [the upper two panels in (a)] and $m=2$
[the lower two panels in (a)]. The dependence of $P$ on $E$ are
shown in (b) and (c) for $m=1$ and $m=2$, respectively. The total
power $P|_{E\rightarrow\infty}\rightarrow2.755$ for $m=1$ and
$P|_{E\rightarrow\infty}\rightarrow5.447$ for $m=2$. The red line
denotes the value of $P_{cr}$ by (\ref{P_cr1}). \label{fig2}}
\end{figure}

In Fig.~\ref{fig2}, we solve Eq.~(\ref{stationary}) by using a
shooting method to construct stationary modes for both Dirichlet and
Neumann boundary conditions and find authentic law for $P(E)$. The
stationary solutions $u(x)$ for $d=2$ are shown. The initial
conditions are taken by the solutions (\ref{linsol-d}) and
(\ref{linsol-n}). We see that the family of solutions with $m=1$
bifurcates from zero at $E=E_1=-1$ while the family of solutions
with $m=2$ bifurcates from zero at $E=E_2=-4$. Close to the
bifurcation points the $P-E$ curves follow the relation
(\ref{E_P2}). $P(E)$ is monotonically increasing with the growth of
$E$ while $d P(E)/d E|_{E\rightarrow\infty}\rightarrow 0$.
%
%===========================fig2===============================%
%===========================fig2===============================%
%

The linear stability of stationary solutions can be numerically
computed by solving the linearized eigenvalue problem. Particularly,
we assume that
\be \psi(x,t)=[u(x)+(\alpha+i\beta)]e^{iEt}, \ee
where $\alpha$, $\beta$ are small perturbations which are
proportional to $\exp(-i\lambda t)$. The coupled eigenvalue problem
then reads
\be L_1B=-i\lambda A,\quad L_2A=i\lambda B, \ee
where $L_1=-\partial_{xx}+E-u(x)^{2d}$ and
$L_2=-\partial_{xx}+E-(2d+1)u(x)^{2d}$. The growth rate is defined
as ${\rm max}[{\rm Im}(\lambda)]$ at which an unstable solution will
grow. We find that the family of solutions with $m=1$ is always
stable against linear perturbations whereas only a very narrow
stability region close to the bifurcation point exists for the
family of solutions with $m=2,3,4\cdots$.

In the case of infinite domain, Eq. (\ref{stationary}) admits the
localized soliton solution
\be \label{sol1} u=[(d+1)E]^{1/2d}{\rm
sech}^{1/d}\left[d\sqrt{E}\left(x-\frac{\pi}{2}\right)\right], \ee
where the solution is centered in the center of the domain
$x=\pi/2$. The total power for $d=2$ is a constant
\be \label{P_cr1} P_{cr}=\int|u|^2dx=\frac{\pi}{4}\sqrt{12}=2.72,
\ee
i.e. it is independent on $E$. The solutions with power $P<P_{cr}$
disperse during the propagation, whereas if $P>P_{cr}$ the solutions
collapse. A similar analysis can be carried out for the case of
$d=3$ leading to
\be \label{P_cr2}
P_{cr}=\int|u|^2dx=\frac{2^{5/3}\pi^{1/2}\Gamma(7/6)}{3^{1/2}\Gamma(2/3)}\frac{1}{E^{1/6}},
\ee
which is expressed in terms of $\Gamma$ functions and depends on
$E$. Although the soliton solutions do not satisfy the NLS equation
on a bounded domain with specific boundary conditions, they are
particularly useful as limiting cases of solutions.

The solution (\ref{sol1}) can be considered as good approximations
for the solutions satisfying both Dirichlet and Neumann boundary
conditions in the limit $E\rightarrow\infty$. This is because in
this latter limit the maximum intensity increases as $E^{1/2}$ while
the pulse width decreases as $E^{1/2}$. The narrowing of the pulse
makes the soliton tails, as well as their derivatives, are almost
zero on the boundaries. The critical power for collapse $P_{cl}$
does not change with the growth of $E$.

\section{The numerical simulations}

To understand the dynamics of the Fourier modes, we relied heavily
on numerical solutions of the NLS equation on the interval
$[0,\pi]$. The equation was solved using the split-step Fourier
method where the linear part is advanced using the sine or cosine
Fourier transforms respectively for Dirichlet and Neumann boundary
conditions. The details of the numerical implementation are given in
the Appendix A.

We first consider Dirichlet boundary condition. The solution can be
expanded in sines as
\be\label{cmt} \psi(t,x)= \sum_{m=1}^\infty c_m(t) \sin mx. \ee
For $d=1$,  there is no collapse for Eq. (\ref{nls}). As expected
the evolution of a sine initial condition gives rise to a cascade of
modes. For $c_1(0)= 2$ and $c_{j\neq1}(0)= 0$ we observe a cascade
to $c_3$ and $c_5$ with maximum amplitudes ${\rm max}(c_3)= 0.7$ and
${\rm max}(c_5)=0.15$ with the other modes being insignificant. For
$c_3(0)= 2$ and $c_{j\neq3}(0)= 0$ we get almost no cascade.

\begin{figure}
\centering
\includegraphics[scale=0.35]{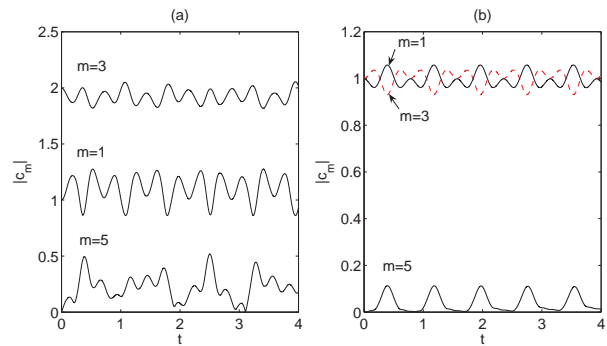}
\caption{\footnotesize  Time evolution for a cubic nonlinearity and Dirichlet
boundary conditions of the
sine mode amplitudes $c_m(t),~m=1,3,5$ starting from two different
initial conditions $\psi(0,x)=\sin(x)+2\sin(3x)$ (left panel)
and $\psi(0,x)=\sin(x)+\sin(3x)$ (right panel). \label{fig3}}
\end{figure}

Now let us compare the outcomes for $d=1$ with the initial
conditions $\psi(0,x)=\sin(x)+2\sin(3x)$ and
$\psi(0,x)=\sin(x)+\sin(3x)$. The time evolution of the mode
amplitudes are shown in Fig.~\ref{fig3}. The mode amplitudes
fluctuate in a fairly narrow range around an average value. This
range decreases even more for smaller initial amplitudes as shown in
the right panel. We will explain these effects in the next section.
%
%===========================fig3===============================%
%===========================fig3===============================%
%

\begin{figure}
\centering
\includegraphics[scale=0.35]{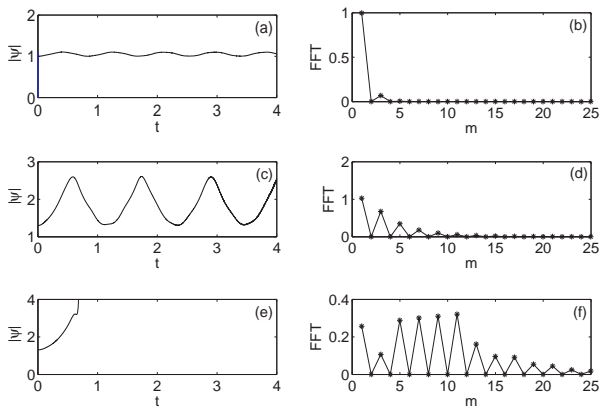}
\caption{\footnotesize Left panels: time evolution of the maximum of
$|\psi|$ for a quintic nonlinearity and Dirichlet boundary conditions
for three different initial conditions,
from the top to the bottom, 
$\psi(0,x)=1.0\sin(x)$ ($P=1.57$),
$\psi(0,x)=1.3\sin(x)$ ($P=2.65$) and $\psi(0,x)=1.31\sin(x)$
($P=2.70$). The right panels show the corresponding Fourier spectra
at time $t=4$. Notice the collapse occurring at $t\approx 0.7$ in
the last row. \label{fig4}}
\end{figure}

Let us now turn to the quintic case ($d=2$) with Dirichlet boundary
condition. In Fig.~\ref{fig4}, we show the time evolution of
solution (\ref{linsol-d}) with $m=1$ and different values of
amplitude $A$. No collapse is observed when $A$ ($P$) is small
whereas collapse occurs when $A> 1.3$ ($P> 2.65$). A collapsing
solution is shown on the last row of Fig.~\ref{fig4} for $A=1.31$.
The right panels of Fig.~\ref{fig4} show
the Fourier spectra. We notice that only the odd modes are
excited. Actually, we can explain that the cascade of Fourier modes
for Eq. (\ref{nls}) starting with a particular mode 
$q$ ($q = 0, 1, 2,\cdots$) is always restricted
to the modes $q(2n-1)$ ($n = 1, 2, 3,\cdots$) irrespective of the
boundary conditions.
In other
words, we can expand the solutions of Eq. (\ref{nls}) as
\be \label{exp-cond} \psi(t,x)=\sum_{n=1}^{\infty} c_{q(2n-1)} \sin[q(2n-1) x]
\ee
for the initial condition $\psi(0,x)=\sin(qx)$. For the Neumann
boundary conditions the sines should be
changed to cosines. Details are given in Appendix B.
In Fig.~\ref{fig5}, we show the recurrence of the solutions and
spectrum cascade at different times when no collapse occurs. We will
explain this phenomenon in the next section. We also notice that for
large propagation constant, the solutions appears in the form of the
hyperbolic secant function. 

In Fig.~\ref{fig6}, we show the time evolution of the Fourier
amplitudes for a initial condition (\ref{linsol-d}) with $m=1$ and
increasing amplitude $A$. Each panel calculated for a single
amplitude corresponds to different regions of similar behaviors for the 
modes.
When $0<A\leq0.5$, there is only the mode $m=1$. The other modes are
insignificant. When $0.5<A\leq1.0$, we observe two modes $m=1$ and
3. When $1.0<A\leq1.2$, we observe three modes $m=1$, 3, 5. When
$1.2<A\leq1.3$, four modes $m=1$, 3, 5, and 7 are observed. The
larger the amplitude of the initial condition, the more modes are 
excited. When $A>1.3$ collapse occurs. This is a much faster mechanism
than the recurrence. The energy travels very suddenly from the low
frequency modes to the higher frequency modes.
%
%===========================fig4===============================%
%===========================fig4===============================%
%
%
%===========================fig5===============================%
\begin{figure}
\centering
\includegraphics[scale=0.33]{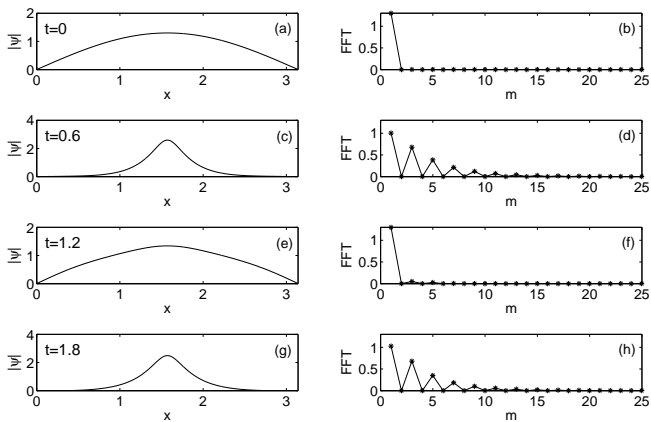}
\caption{\footnotesize Snapshots of the wave profiles and Fourier
spectra for a quintic nonlinearity and Dirichlet boundary condition.
The times shown are $t=0,~ 0.6,~ 1.2$ and 1.8 respectively in panels
a, b, c and d. The initial condition is $\psi(0)=1.3\sin(x)$
($P=2.65$). The recurrence of the wave profile and the spectrum
cascade is evident. The profiles shown in (c) and (g) follow the 
hyperbolic secant function (\ref{sol1}). \label{fig5}}
\end{figure}
%===========================fig5===============================%
%
%
%===========================fig6===============================%
\begin{figure}
\centering
\includegraphics[scale=0.33]{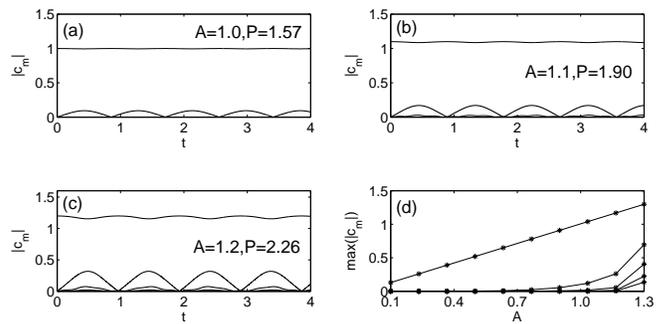}
\caption{\footnotesize Time evolution of the Fourier amplitudes for
quintic nonlinearity and Dirichlet boundary condition for different
initial conditions. In panels a, b and c, $c_1(0)=A=1., ~ 1.1$ and
1.2, respectively. The panel d shows ${\rm max}(|c_m|)$ for $m=1,~
3,~ 5,~ 7$ and 9 as a function of the initial amplitude $A$. }
\label{fig6}
\end{figure}
%===========================fig6===============================%
%
For the Neumann boundary conditions, the zero mode plays a special
role. For example, it does not give rise to a spectrum cascade. 
We will see below how it couples to the other modes.
Apart from this, the simulation results for the Neumann boundary conditions
are similar to the ones for the Dirichlet boundary conditions.

\section{Evolution of the resonant Fourier modes }

In the interval of existence of solutions, one can use the Fourier
sine or cosine series depending whether we have Dirichlet or Neumann
boundary conditions. In both cases we have a complete basis for
functions satisfying the boundary conditions. 
We will first consider the Dirichlet boundary condition and
examine cubic and quintic nonlinearities.
The results that we will obtain for these are very similar
 for the Neumann boundary
condition unless the zero mode is involved. This specific case will 
be addressed at the end of the section.

For the Dirichlet boundary conditions, we expand $\psi$ in a sine Fourier series
\be \label{exp-d} \psi(x,t)=\sum_{m=1}^{\infty}c_m(t)\sin(mx), \ee
where the $c_m$ are given by
\be c_m(t)=\frac{2}{\pi}\int_0^{\pi}\psi(x,t)\sin(nx)dx. \ee
We find that if collapse occurs, it happens due to the divergence of
the Fourier series. Indeed, the conservation of the power implies
for any $m'$
\be |c_{m'}(t)|^2\leq \sum_{m=1}^{\infty}|c_{m}(t)|^2={2\over \pi} P \equiv {\cal P}, \ee
because of Parseval's relation,
$$\int_0^\pi |\psi|^2 dx = {\pi\over 2} {\cal P} .$$
Therefore the Fourier coefficients are bounded and the only source of
collapse is the divergence of the series.

Substituting the expansion (\ref{exp-d}) into Eq. (\ref{nls}), we
obtain the coupled equations of Fourier amplitudes. For 
the quintic nonlinearity ($d=2$) we get 
\bea \label{fe-d}
& &
i\dot{c}_q-q^2c_q+\frac{2}{\pi}\sum_{k,l,m,n,p}c_kc_lc_mc_n^{\ast}c_p^{\ast}\langle
klmnp|q\rangle\nonumber\\
& &=0, \eea
where
\bea \langle
klmnp|q\rangle&\equiv&\int_0^\pi\sin(kx)\sin(lx)\sin(mx)\sin(nx)\nonumber\\
& &\times\sin(px)\sin(qx)dx. \nonumber\eea
Eqs. (\ref{fe-d}) can be simplified by the transformation
$c_q=a_qe^{-i q^2t}$ yielding
\bea \label{fe-d1}
& &
i\dot{a}_q+\frac{2}{\pi}\sum_{k,l,m,n,p}a_ka_la_ma_n^{\ast}a_p^{\ast}\langle
klmnp|q\rangle\nonumber\\
& & \quad \times e^{-i (k^2+l^2+m^2-n^2-p^2-q^2)t}=0. \eea
Similarly, the coupled equations of Fourier amplitudes for the case
of septic nonlinearity ($d=3$) read
\bea \label{fe-d2}
& &
i\dot{a}_s+\frac{2}{\pi}\sum_{k,l,m,n,p,q,r}a_ka_la_ma_na_p^{\ast}a_q^{\ast}a_r^{\ast}\langle
klmnpqr|s\rangle\nonumber\\
& & \quad \times e^{-i(k^2+l^2+m^2+n^2-p^2-q^2-r^2-s^2)t}=0. \eea
Note that in Eqs. (\ref{fe-d1}) and (\ref{fe-d2}) most terms are 
rapidly rotating and average out to zero. Only the ones such that
$k^2+l^2+m^2-n^2-p^2-q^2=0$ for (\ref{fe-d1}) 
( $k^2+l^2+m^2+n^2-p^2-q^2-r^2-s^2=0$  for (\ref{fe-d2})) i.e., 
the resonant terms will contribute to the long term dynamics of
${a}_q$.

A detailed study can be carried out of the dynamics of the Fourier
coefficients for different types of nonlinearities. We have used the
Maple software \cite{Maple} to identify all the resonant terms in the equations
for the mode amplitudes. For the cubic nonlinearity $(d=1)$ the
amplitude equation equivalent to (\ref{fe-d1}) reads
\bea \label{fe-d0}
& &
i\dot{a}_j+\frac{2}{\pi}\sum_{k,l,m}a_ka_la_m^{\ast}\langle
klm|j\rangle\nonumber\\
& & \quad \times e^{-i (k^2+l^2-m^2-j^2)t}=0. \eea
Taking into account the resonance condition,
Eq. (\ref{fe-d0}) turns into the following equations
\be \label{fe-d3} i\dot{a}_j+ a_j\left(
{\cal P} -\frac{|a_j|^2}{4}\right)=0,
\quad(j=1,\,2,\cdots \infty) \ee
where ${\cal P} =\sum_{j=1}^{\infty}|a_j|^2$ is conserved. 
Eq. (\ref{fe-d3}) admits the solution
\be a_j=|a_j|e^{i({\cal P}-|a_j|^2/4)t}. \ee
An obvious implication is that $d/dt(|a_j|^2)= 0$ so that 
there is no transfer of energy from one mode to
another. This is what we have seen in the numerical results in Fig.
\ref{fig3}. Over a short time the modes oscillate in a periodic
fashion, however their average over a long time is constant. 
As expected no collapse will occur in the
cubic NLS equation irrespective of the initial total power. In fact the
integrability of the NLS equation on the whole line, related to the existence
of a Lax pair, has been shown to carry over to the case of a finite domain
with Dirichlet boundary conditions \cite{fokas_its04}.

The case of the quintic nonlinearity $(d=2)$ is more complicated.
For simplicity, we consider a solution consisting of three modes,
i.e., $m=1,\,3$ and $5$,
\bea \label{exs} \psi(t,x)&=&a_1e^{-i t}\cos(x)+a_3e^{-i 9 t}\cos(3x)\nonumber\\
&+&a_5e^{-i 25 t}\cos(5x). \eea
Then Eq. (\ref{fe-d1}) gives rise to the following coupled resonant amplitude equations
\bes \label{fe-d4} \bea
& & i\dot{a}_1+
a_1\left[\frac{9}{4}|a_1|^2{\cal P}-\frac{13}{8}|a_1|^4+3|a_3|^2|a_5|^2\right.\nonumber\\
& & \quad
\left.+\frac{9}{8}\left(|a_3|^4+|a_5|^4\right)\right]-\frac{3}{8}
a_1^{\ast}a_3^3a_5^{\ast}=0,\\
& & i\dot{a}_3+
a_3\left[\frac{9}{4}|a_3|^2{\cal P}-\frac{13}{8}|a_3|^4+3|a_1|^2|a_5|^2\right.\nonumber\\
& & \quad
\left.+\frac{9}{8}\left(|a_1|^4+|a_5|^4\right)\right]-\frac{9}{16}
a_1^2a_3^{\ast 2}a_5=0,\\
& & i\dot{a}_5+
a_5\left[\frac{9}{4}|a_5|^2{\cal P}-\frac{13}{8}|a_5|^4+3|a_1|^2|a_3|^2\right.\nonumber\\
& & \quad
\left.+\frac{9}{8}\left(|a_1|^4+|a_3|^4\right)\right]-\frac{3}{16}
a_1^{\ast 2}a_3^3=0, \eea \ees
where ${\cal P}=|a_1|^2+|a_3|^2+|a_5|^2$. It is easy to check that Eqs.
(\ref{fe-d4}) satisfy the condition
$\frac{d}{dt}(|a_1|^2+|a_3|^2+|a_5|^2)=0$, i.e., ${\cal P}$ is a conserved
quantity. The last terms in equations represent the mixing between
the three modes and therefore the intensity of each mode is not
conserved as in the case of $d=1$. At this point note that if we 
had included $a_7$ in the
description we would have had the additional resonant terms 
$$ |a_3|^2 a_5^2 a_7,~~
a_5^2 |a_7|^2 a_7^\ast,~~
a_5^2 |a_5|^2 a_7^\ast.$$ If $|a_7| << 1$ then their contribution
would be very small. Another point is that if we had included the even
modes in the description we would have obtained the extra resonant term
$a_2^2 a_4^{\ast 2} a_5$ in the equation for $a_1$. This indicates that
the modes 2 and 4 couple through mode 5.

Eqs. (\ref{fe-d4}) can be written into a more compact form by
defining $I_j=|a_j|^2$ ($j=1,\,3,\,5$). Here, $I_j$ represent the
intensity of each mode satisfying ${\cal P}=I_1+I_3+I_5$. Then, we obtain
\bes \label{fe-d5} \bea
& & \dot{I}_1=\frac{3}{4}
I_1I_3^{3/2}I_5^{1/2}\sin\theta,\\
& & \dot{I}_3=-\frac{9}{8}
I_1I_3^{3/2}I_5^{1/2}\sin\theta,\\
& & \dot{I}_5=\frac{3}{8} I_1I_3^{3/2}I_5^{1/2}\sin\theta, \eea \ees
where $\theta=-2\theta_1+3\theta_3-\theta_5$ with $\theta_j={\rm
arg}\,a_j$. We notice that the driving terms on the right hand side
of Eqs. (\ref{fe-d5}) are proportional to the intensities of the
modes and their phases. In addition, we have the constraints
\bes \label{constraint} \bea
& & \frac{4}{3}I_1+\frac{8}{9}I_3=\mu_1,\\
& & \frac{8}{9}I_3+\frac{8}{3}I_5=\mu_2,\\
& & \frac{4}{3}I_1-\frac{8}{3}I_5=\mu_1-\mu_2=\mu_3,\eea \ees
where $\mu_i,~(i=1-3)$ are constants of the motion. Using the above
relations, the dynamics can be reduced to the equations for $I_1$
and $\theta$. They are
\bes \label{fe-d6} \bea & & \dot{I}_1=\frac{3}{4}
I_1I_3^{3/2}I_5^{1/2}\sin\theta,\\
& & \dot{\theta}= \frac{1}{4}(13 I_1^2 +3 I_3^2 +7 I_5^2 +21 I_1 I_3 +27 I_1 I_5 +12 I_3 I_5) \nonumber\\
& & ~ + \frac{\cos\theta}{8}[6 I_1^{1/2}I_3^{3/2}I_5^{1/2} -27
I_1I_3I_5^{1/2} \nonumber\\
& & ~ -3I_1I_3^{3/2}I_5^{-1/2}] , \eea \ees
together with the constraints (\ref{constraint}).

\begin{figure}
\centering
\includegraphics[scale=0.35]{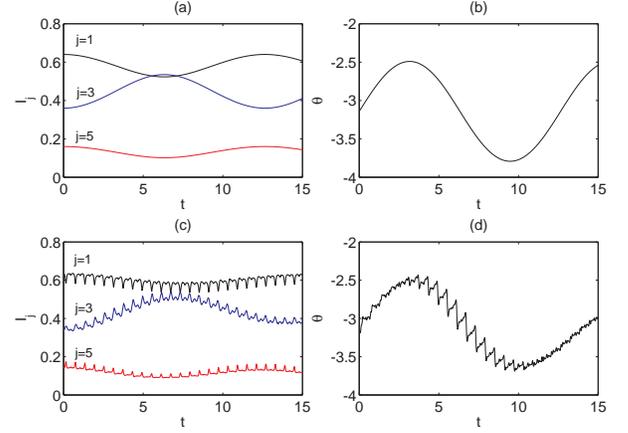}
\caption{\footnotesize Comparison between the solution of the
reduced model and the numerical solution of the NLS for quintic
nonlinearity and Dirichlet boundary condition. The top left (right)
panel shows the time evolution $I_1(t)$, $I_3(t)$, and $I_5(t)$
($\theta(t)$) for the reduced model (\ref{fe-d6}). The bottom left
(right) panel shows the corresponding evolutions for the NLS. The
initial conditions are taken as $I_1=0.64$, $I_3=0.36$ and
$I_5=0.16$ and $\theta=-\pi$. \label{fig7}}
\end{figure}

In order to check the equations for the resonant Fourier modes, we
compare the solutions of the reduced equations and
the NLS. In Fig.~\ref{fig7} we show the time evolution of the solutions
of Eqs. (\ref{fe-d5}) and (\ref{nls}) by using the Runge-Kutta
method and split-step Fourier method, respectively. The initial
conditions are the same. As expected the amplitudes of the Fourier
modes for (\ref{nls}) present fast periodic oscillations. However
over a long time interval the solutions of the reduced equations
and the full partial differential equation  
match well, supporting the validity of the reduced model.

At this point, let us consider the Neumann boundary conditions for which there
is the additional zero mode. For the cubic nonlinearity $d=1$, the evolution
of the Fourier modes follows (\ref{fe-d3}) so that there is no resonant energy
exchange between the modes $\dot I_j =0, j=0,1,\dots$.
For the quintic nonlinearity $d=2$, the situation is more interesting. 
Assuming a solution containing the three modes $i=1,3,5$, we obtain 
evolution equations identical to (\ref{fe-d4}) except that the signs
of the resonant terms are reversed. We then obtain the same final equations
(\ref{fe-d6}) except that the evolution of the phase $\theta$ is reversed.

Assuming a solution containing the three first modes $i=0,1,2$ , we obtain evolution
equations of the form (\ref{fe-d4}) but with no terms outside the brackets. This
means that again, no resonant transfer of energy exists between modes. If the third mode
is added to the expansion, new terms appear outside the brackets. The evolution
equations are
\bes \label{fe-d4a} \bea
& & i\dot{a}_0+ a_0\left[\dots \right]+ \frac{3}{4} a_0^{\ast}a_1 a_2^2 a_3^{\ast}=0,\\
& & i\dot{a}_1+ a_1\left[\dots \right]+ \frac{3}{4} a_0^{2} a_2^{\ast 2} a_3=0,\\
& & i\dot{a}_2+ a_2\left[\dots \right]+ \frac{3}{2} a_0^{2}a_1^{\ast} a_2^{\ast} a_3=0,\\
& & i\dot{a}_3+ a_3\left[\dots \right]+ \frac{3}{4} a_0^{\ast 2}a_1 a_2^2 =0,
 \eea \ees
where the $\dots$ terms in the brackets are all real. Following a 
similar procedure
as above, the modal energies $I_j = |a_j|^2$ evolve as
\bes \label{fe-d5a} \bea
& & \dot{I}_0=\frac{3}{2} I_0 I_1^{1/2}I_2 I_3^{1/2}\sin\theta,\\
& & \dot{I}_1=\frac{3}{2} I_0 I_1^{1/2}I_2 I_3^{1/2}\sin\theta,\\
& & \dot{I}_2=3 I_0 I_1^{1/2}I_2 I_3^{1/2}\sin\theta,\\
& & \dot{I}_3=\frac{3}{2} I_0 I_1^{1/2}I_2 I_3^{1/2}\sin\theta,
\eea \ees
where $\theta = -2 \theta_0+ \theta_1 + 2 \theta_2 -\theta_3$. As above one can
then reduce the problem to two equations, one for $I_0$ and one for $\theta$.
We do not write these equations because they are cumbersome. The interesting fact
is that one needs the 4 modes 0-3 present in order to see this resonant transfer of
energy. If one of them is missing there is no energy transfer. This particular
feature of Neumann boundary conditions changes the route for collapse for the
Dirichlet case and the Neumann case.

%
%===========================fig7===============================%
%===========================fig7===============================%
%

\section{Filtering Fourier modes prevents collapse}

The projection of the nonlinear Schr\"{o}dinger equation on a 
a finite number of Fourier modes will yield amplitude equations
that are well behaved and do not exhibit collapse. As we have
seen collapse is related to a sudden energy flow to high frequencies.
Thus, one can arrest the collapse by filtering the high Fourier
modes and therefore preventing the sudden energy flow to high frequencies.
Physically this can be done by
introducing a nonlocal absorption in the model (\ref{nls}). This
conclusion is also available for the models with septic nonlinearity
and Neumann boundary condition.

\begin{figure}
\centering
\includegraphics[scale=0.35]{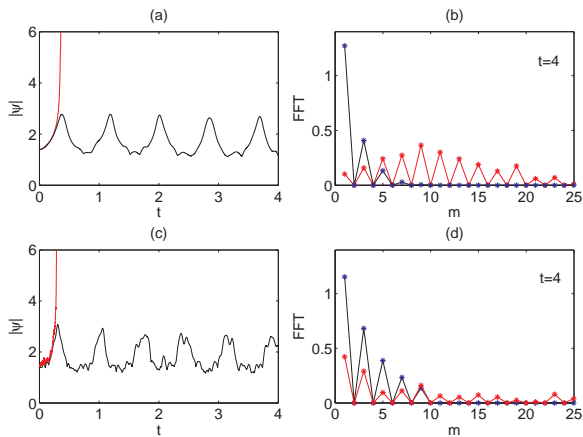}
\caption{\footnotesize Left panels: Time evolution of the maximum of
$|\psi|$ for quintic nonlinearity with different initial conditions.
Right panels: The corresponding Fourier spectra at $t=4$. The
initial conditions are taken as $\psi(0,x)=1.4\sin(x)$ ($P(0)=3.08$)
in the first row and $\psi(0,x)=1.4\,{\rm sech}(x-\frac{\pi}{2})$
($P(0)=3.56$) in the second row. We compare the results with and
without filtering the high Fourier modes designated by the black and
red lines, respectively. The collapse is efficiently prevented by
filtering in both cases. \label{fig8}}
\end{figure}

In Fig.~\ref{fig8} we show the time evolution of solution with or
without filtering high Fourier modes. The initial conditions are
taken as a sine wave (the first row) and a hyperbolic secant pulse
(the second row). In both cases, the collapse is efficiently
prevented by filtering. Because of the absorption of the higher
modes, there is a small loss (less than 10 percent) of the total
power. This can be decreased by including more lower frequency modes.
%
%===========================fig8===============================%
%===========================fig8===============================%
%

\section{Conclusion}

We have analyzed and solved numerically the one dimensional
nonlinear Schr\"{o}dinger equation on a finite interval with
Dirichlet or Neumann boundary conditions.
A preliminary analysis reveals that $H<0$ is sufficient 
for collapse in the Dirichlet case but not for the Neumann case.
The bound states have been computed. The first non trivial one 
corresponding in the linear limit to $\sin x$ (resp. $\cos x$) for 
Dirichlet (resp. Neumann) boundary conditions 
is always linearly stable as opposed to the higher order 
modes whose window of stability is very small
and reduces as the order is increased.

We have solved the partial differential equation for the cubic 
and quintic nonlinearities. In the cubic case there is no
resonant transfer of energy between Fourier modes, while it
is present in the quintic case. Identifying resonant terms in the
evolution equations of the Fourier modes, we have written 
reduced systems. Their evolution is in excellent agreement with 
the solutions of the NLS, even close to collapse. 
For the Neumann boundary conditions the Goldstone
mode plays a particular role as it couples the modes $m=1,2,3$.
For the Dirichlet case, there is no coupling between the first three
modes.
Finally note that this model reduction can be extended to higher 
dimensions and other systems like a cylindrical waveguide. The machinery 
would be more complicated but the overall method remains the same.

\acknowledgments

The research of JGC and CH was supported by a joint France-Portugal Pessoa 
agreement. 
The research of CH was supported by the
Funda\c{c}\~ao para a Ci\^encia e a Tecnologia (FCT) under Grant No.
SFRH/BPD/36385/2007 and Est\'{i}mulo \`{a} Investiga\c{c}\~{a}o 2009
de Funda\c{c}\~{a}o Calouste Gulbenkian. The authors thank Vladimir Konotop
for very helpful discussions.

\appendix

\section{Numerical procedure for solving the 1D NLS (\ref{nls})}

The 1D nonlinear Schr\"{o}dinger equation (\ref{nls}) with Dirichlet or
Neumann boundary conditions is solved as usual by splitting the
linear and nonlinear part of the operator. The linear part
\be\label{linstep} iu_t +u_{xx}=0,\ee is such that
$${\hat u}(dt) = e^{-ik^2 dt} {\hat u}(0),$$
where ${\hat u}$ is the sine or cosine Fourier transform of $u$, to
satisfy the boundary conditions. The solution of the nonlinear
stepping is the standard one
$$u(2dt)= e^{i |u(dt)|^{2d}dt}u(dt).$$
The numerical implementation is done in Matlab \cite{Matlab} and the
solution is evaluated at discrete points $u_n, n=1,\dots N$. The
sine and cosine Fourier transforms for the linear step
(\ref{linstep}) are then done using the discrete sine and cosine
Fourier transforms.

For the Dirichlet boundary conditions, we use the discrete sine
Fourier transform \be\label{dst} {\hat u(k)} = \sum_{n=1}^N u(n)
\sin({\pi k n  \over N+1}),~~k=1,\dots N,\ee and inverse discrete
sine Fourier transform \be\label{idst} {u(n)} = \sum_{k=1}^N {\hat
u}(k) sin({\pi k n  \over N+1}),~~n=1,\dots N.\ee

The Neumann boundary conditions are trickier to implement because
one needs to use the discrete cosine Fourier transform
\be\label{dct} {\hat u(k)} = w(k) \sum_{n=1}^N u(n) \cos({\pi (2 n
-1)(k-1) \over 2 N}),~~k=1,\dots N,\ee with $$w(1)=
1/\sqrt{N},~~w(k)= \sqrt{2\over N},~~2\le k\le N$$. The inverse
discrete cosine Fourier transform is \be\label{idct} {u(n)} =
\sum_{k=1}^N w(k) {\hat u}(k) \cos({\pi (2 n -1)(k-1) \over 2
N}),~~n=1,\dots N.\ee

The number of discretisation points was chosen to be $N=2^{11}-1$ or
$N=2^{12}-1$ with a step $dt=10^{-4}$. The L$_2$ norm was checked
during the computation and it is conserved up to $10^{-10}$ in
absolute value.

\section{Parity argument for the mode cascade}

In order to prove Eq. (\ref{exp-cond}), let us consider the case of
$d=1$, the cases of $d=2,~ 3$ can be proved in a similar way. Eq.
(\ref{nls}) with $d=1$ can be written into the form
\bea \label{ie} & &
\psi(t+dt,x)=\psi(t,x)+i\left[\frac{1}{2}\frac{\partial^2
\psi(t,x)}{\partial
x^2}\right.\nonumber\\
& &\left.+|\psi(t,x)|^{2}\psi(t,x)\right]dt. \eea
Now we assume that the initial condition is taken as
$\psi(0,x)=\cos(qx)=\frac{1}{2}(e^{iqx}+e^{-iqx})$. Substituting the
initial condition into Eq. (\ref{ie}), we have
\bea & & \psi(dt,x)=
\frac{1}{2}(e^{iqx}+e^{-iqx})+i\left[-\frac{1}{4}(e^{iqx}+e^{-iqx})\right.\nonumber\\
&
&\left.+\frac{1}{8}(e^{3iqx}+3e^{iqx}+3e^{-iqx}+e^{-3iqx})\right]dt.
\eea
As we see, after a short time interval $dt$, the solution
$\psi(dt,x)$ can be expressed as
$\psi(dt,x)=c_q\cos(qx)+c_{3q}\cos(3qx)$. We can repeat this process
and obtain that $\psi(t,x)=\sum_{m=(2n-1)q} c_m\cos(mx)$
($n=1,2,\cdots$). In other words for $d=1$ we immediately obtain
modes 1, 3 and 5 while for $d=2$ we immediately obtain modes 1, 3, 5
and 7, etc. Finally, notice that for Neumann boundary condition
there is no cascade to higher Fourier modes starting with the $m=0$
mode.

\section{Evolution of the resonant Fourier modes for the Neumann boundary conditions}

For the Neumann case, we expand $\psi$ in a cosine Fourier series
\be \label{exp-n} \psi(x,t)=\sum_{m=0}^{\infty}c_m(t)\cos(mx), \ee
where the $c_m$ are given by
\bes
\bea c_0(t)&=&\frac{1}{\pi}\int_0^{\pi}\psi(x,t)dx,\\
c_n(t)&=&\frac{2}{\pi}\int_0^{\pi}\psi(x,t)\cos(nx)dx \quad(n\neq0).
\eea \ees

Substituting the expansion (\ref{exp-n}) into Eq. (\ref{nls}), we
obtain the coupled equations of Fourier amplitudes for the case of
quintic nonlinearity ($d=2$)
\bes \label{fe-n} \bea
& &
i\dot{c}_0+\frac{1}{\pi}\sum_{k,l,m,n,p}c_kc_lc_mc_n^{\ast}c_p^{\ast}\langle
klmnp|0\rangle=0,\\
& &
i\dot{c}_q-q^2c_q+\frac{2}{\pi}\sum_{k,l,m,n,p}c_kc_lc_mc_n^{\ast}c_p^{\ast}\langle
klmnp|q\rangle\nonumber\\
& &=0,\quad (q\neq0) \eea \ees
where
\bea \langle
klmnp|q\rangle&\equiv&\int_0^\pi\cos(kx)\cos(lx)\cos(mx)\cos(nx)\nonumber\\
& &\times\cos(px)\cos(qx)dx. \nonumber\eea
Eqs. (\ref{fe-n}) can be simplified by the transformation
$c_q=a_qe^{-i q^2t}$ yielding
\bea \label{fe-n1}
& &
i\dot{a}_q+\frac{\sigma_q}{\pi}\sum_{k,l,m,n,p}a_ka_la_ma_n^{\ast}a_p^{\ast}\langle
klmnp|q\rangle\nonumber\\
& & \quad \times e^{-i (k^2+l^2+m^2-n^2-p^2-q^2)t}=0, \eea
where $\sigma_q=1$ for $q=0$ and $\sigma_q=2$ for $q\neq0$.
Similarly, the coupled equations of Fourier amplitudes for the case
of septic nonlinearity ($d=3$) read
\bea \label{fe-n2}
& &
i\dot{a}_s+\frac{\sigma_s}{\pi}\sum_{k,l,m,n,p,q}a_ka_la_ma_na_p^{\ast}a_q^{\ast}a_r^{\ast}\langle
klmnpqr|s\rangle\nonumber\\
& & \quad \times e^{-i\frac{1}{2}
(k^2+l^2+m^2+n^2-p^2-q^2-r^2-s^2)t}=0, \eea
where $\sigma_s=1$ for $s=0$ and $\sigma_s=2$ for $s\neq0$.

For simplicity, we consider the solution (\ref{exs}). Then, Eq.
(\ref{fe-n1}) turns into the following coupled equations
\bes \label{fe-n3} \bea
& & i\dot{a}_1+
a_1\left[\frac{9}{4}|a_1|^2P-\frac{13}{8}|a_1|^4+3|a_3|^2|a_5|^2\right.\nonumber\\
& & \quad
\left.+\frac{9}{8}\left(|a_3|^4+|a_5|^4\right)\right]+\frac{3}{8}
a_1^{\ast}a_3^3a_5^{\ast}=0,\\
& & i\dot{a}_3+
a_3\left[\frac{9}{4}|a_3|^2P-\frac{13}{8}|a_3|^4+3|a_1|^2|a_5|^2\right.\nonumber\\
& & \quad
\left.+\frac{9}{8}\left(|a_1|^4+|a_5|^4\right)\right]+\frac{9}{16}
a_1^2a_3^{\ast 2}a_5=0,\\
& & i\dot{a}_5+
a_5\left[\frac{9}{4}|a_5|^2P-\frac{13}{8}|a_5|^4+3|a_1|^2|a_3|^2\right.\nonumber\\
& & \quad
\left.+\frac{9}{8}\left(|a_1|^4+|a_3|^4\right)\right]+\frac{3}{16}
a_1^{\ast 2}a_3^3=0, \eea \ees
where ${\cal P}=|a_1|^2+|a_3|^2+|a_5|^2$. These are exactly the same
as (\ref{fe-d4}) except that the terms outside the brackets have
the opposite signs.

As done above, eqs. (\ref{fe-n3}) can be written into a more compact form by
defining $I_j=|a_j|^2$ ($j=1,~3,~5$), i.e.
\bes \label{fe-n4} \bea
& & \dot{I}_1=-\frac{3}{4}
I_1I_3^{3/2}I_5^{1/2}\sin\theta,\\
& & \dot{I}_3=\frac{9}{8}
I_1I_3^{3/2}I_5^{1/2}\sin\theta,\\
& & \dot{I}_5=-\frac{3}{8} I_1I_3^{3/2}I_5^{1/2}\sin\theta, \eea
\ees
where $\theta=-2\theta_1+3\theta_3-\theta_5$ with $\theta_j={\rm
arg}\,a_j$. Eq. (\ref{fe-n4}) can be further written into the form
\bes \label{fe-n5} \bea & & \dot{I}_1=-\frac{3}{4}
I_1I_3^{3/2}I_5^{1/2}\sin\theta,\\
& & \dot{\theta}= \frac{1}{4}(13 I_1^2 +3 I_3^2 +7 I_5^2 +21 I_1 I_3 +27 I_1 I_5 +12 I_3 I_5) \nonumber\\
& & ~ - \frac{\cos\theta}{8}[6 I_1^{1/2}I_3^{3/2}I_5^{1/2} -27
I_1I_3I_5^{1/2} \nonumber\\
& & ~ -3I_1I_3^{3/2}I_5^{-1/2}], \eea \ees
together with the constraints (\ref{constraint}).

%%%%%%%%%%%%%%%%%%%%%%%%%%%%%%%%%%%%%%%%%%%%%%%%%%%%%%%%%%%%%%

%%%%%%%%%%%%%%%%%%%%%%%%%%%%%%%%%%%%%%%%%%%%%%%%%%%%%%%%%%%%%%

\end{document}